\documentclass[aps,pre,twocolumn,superscriptaddress,amsmath,amssymb]{revtex4-1}

\usepackage{times}
\usepackage{xcolor}

\usepackage{graphicx}

\usepackage{bm}

\newcommand{\re}{\operatorname{Re}}
\newcommand{\im}{\operatorname{Im}}

\newcommand{\ve}[1]{\mathbf{ #1} }

\renewcommand{\d}{\mathrm{d}}

\newcommand{\F}[1]{(\ref{#1})}

\begin{document}
\title{Synchronization-Desynchronization Transitions in Complex Networks: \\ An Interplay of Distributed Time Delay and Inhibitory Nodes}

\author{Carolin Wille}
\author{Judith Lehnert}
\author{ Eckehard Sch{\"o}ll}
\email{ schoell@physik.tu-berlin.de}
\affiliation{Institut f{\"u}r Theoretische Physik, Technische
Universit{\"a}t Berlin, Hardenbergstr. 36, 10623 Berlin, Germany}

\begin{abstract}
We investigate the combined effects of distributed delay and the balance between excitatory and inhibitory nodes on the stability of synchronous oscillations in a network of coupled Stuart--Landau oscillators. To this end a symmetric network model is proposed for which the stability can be investigated analytically. It is found that beyond a critical inhibition ratio synchronization tends to be unstable. However, increasing distributional widths can counteract this trend leading to multiple resynchronization transitions at relatively high inhibition ratios. The extended applicability of the results is confirmed by numerical studies on asymmetrically perturbed network topologies. All investigations are performed on two distribution types, a uniform distribution and a Gamma distribution.

\end{abstract}
\maketitle

\section{Introduction}
The balance between excitatory and inhibitory elements is of great relevance for processes of self-organization in various neurological structures \cite{WIL72,SHU03,HAI06,VOG09}. Synchronization is one prominent phenomenon that is critically influenced by the balance of excitation and inhibition \cite{BUZ95,ERM98,BRU00,SAL01,BEL08,ERN11,NIS11,LAN12} and is itself an important mechanism involved in processes such diverse as learning and visual perception on the one hand \cite{SIN99b} and the occurrence of Parkinson's disease and epilepsy on the other hand \cite{TAS98,POE01}. Exploring the preconditions for synchronization on a theoretical level can be effectively performed within the framework of complex dynamical networks -- a field that has gained much attention during the last decades \cite{ALB02a,BOC06a}. 

The influence of a particular network architecture can be rich and is widely studied. However, the interaction between the elements of the network can be modeled not only by the strength of the connecting links, but in addition, time-delayed coupling can be incorporated to take into account finite signal transmission times. A single constant delay time, referred to as discrete time-delay, already influences the system's dynamics \cite{LI04} and its synchronization properties \cite{KIN09,CHO09,LEH11,KEA12,SCH13,FLU13} dramatically. Yet, more complex concepts of time-delay have been considered recently to achieve a better correspondence between the model and natural systems \cite{SIP08,EUR05,CAK14}. 

A distribution of delay-times arises naturally, if a model is constructed from experimental data. For instance, the delay distribution in a specific avian neural feedback loop can be well approximated by a Gamma distribution \cite{MEY08}. 
Additionally, time-dependent delay-times can as well be represented by distributed delay, if the delay-time varies rapidly compared to intrinsic time-scales of the system \cite{GJU10,GJU13}. Recent theoretical studies of systems with distributed delay have addressed amplitude death \cite{ATA03,ATA08,KYR11,KYR13} and non-invasive stabilization of periodic orbits \cite{CHO14}.

In this work, we study the effects of distributed time-delay on synchronization in a network model with inhibitory nodes. The stability of synchronization is investigated using the master stability function (MSF) approach \cite{PEC98}. With this technique the effect of the network topology can be separated from the dynamics of the network's constituents, which is described by the master stability function of a complex parameter $\gamma$. The eigenvalues of the coupling matrix representing the network structure then yield the stability for a specific network when they are inserted for the parameter $\gamma$. Thus, distributed delay, which affects the dynamics, and inhibition, which alters the eigenvalue spectrum of the coupling matrix, can both be investigated independently. 

This work extends previous works \cite{LEH11,KEA12} on inhibition-induced desynchronization in time-delayed networks in two ways. First, distributed delay is considered instead of discrete delay. Second, a more realistic network model based on inhibitory nodes instead of inhibitory links is proposed and studied. Such a model might be more appropriate especially in the context of neuroscience, where for the majority of neurons it is established that the same set of neurotransmitters is released at each synapse, a rule which is often referred to as Dale's law. Thus, it is reasonable to assume that in most cases a neuron either excites all its neighbors or inhibits all its neighbors, and not a combination of both.

In contrast to previous work \cite{LEH11,KEA12} where the inhibitory links have been added randomly, we develop a highly symmetric model, which has two main advantages. First, the eigenvalue spectrum of the coupling matrix is real and can be calculated analytically. Second, in networks with randomly added inhibition, the increase of inhibition is accompanied by the increase of asymmetry. Thus, it is not possible to study the pure effect of inhibition, but only a combined effect of asymmetry and inhibition. This difficulty does not arise in the network model proposed here.

As the network's constituents we choose Stuart--Landau oscillators. In contrast to simple phase-oscillators like the Kuramoto oscillator, this paradigmatic non-linear oscillator has an additional radial degree of freedom, which leads to more complex dynamics, i.e., coupled amplitude and phase dynamics. Further, it can be used to describe generally any system close to a Hopf bifurcation, which occurs in a variety of prominent systems used to model e.g., semiconductor lasers \cite{TRO00}, neural systems like the FitzHugh-Nagumo model \cite{FIT61,NAG62} or the Morris-Lecar model \cite{TSU06a}.

This paper is structured as follows. In Section \ref{sec_balance}, we introduce the network model. We derive and characterize the eigenvalue spectrum of its coupling matrix. In Section \ref{sec_sync}, we investigate the existence of synchronous oscillations with respect to the distributed delay parameters. Subsequently, we analyze their stability in Section \ref{sec_stable}. In Section \ref{sec_trans}, we discuss how the interplay of distributed delay and inhibitory nodes influences the stability for the specific network model proposed here. The robustness of our results against asymmetric perturbations of the network topology is investigated numerically in Section \ref{sec_asym}. We conclude our work in Section \ref{sec_conc}.

\section{Balance of Excitation and Inhibition} \label{sec_balance}
The topology of a network is encoded in its coupling matrix $G$. Here, we consider networks without self-coupling $G_{ii}=0$ and normalized row sum 
\begin{equation}
\sum_{j} G_{ij} =1 \;, \label{rowsum}
\end{equation}
which allows for the existence of synchronous solutions. As the stability of the synchronous state depends on the eigenvalue spectrum of $G$, we are interested in the question how the balance between excitatory and inhibitory nodes influences the latter.

It is known that inhibition corresponding to negative entries in the coupling matrix increases the spreading of the eigenvalue distribution in the complex plane \cite{LEH11,KEA12}. For a symmetric network with excitatory nodes only, all eigenvalues are real and located within the interval $[-1,1]$ due to Gershgorin's circle theorem \cite{GER31}. Adding inhibition can cause the existence of eigenvalues with absolute values larger than one. Furthermore, introducing inhibition randomly causes asymmetric coupling matrices that have complex eigenvalues. These two effects are usually intertwined. However, it is desirable to investigate the effects of inhibition independently from the effect of asymmetry.

 To this end, we introduce a simple network model with excitatory and inhibitory nodes, whose coupling matrix has a real eigenvalue spectrum that can be calculated analytically. Thus, it is capable of providing clear statements about the influence of inhibitory nodes that underline similar numerical results found previously \cite{KEA12}.

The network model we propose is composed of two regular rings. One consists of $n$ excitatory nodes coupled to their $k$ nearest neighbors with equal weight $a$ of the links. The other consists of $m$ inhibitory nodes coupled to their $l$ nearest neighbors with links of weight $-a$. 
All excitatory nodes are coupled to all inhibitory nodes with equal link strength $b$. Likewise, all inhibitory nodes are coupled to all excitatory ones with link strength $-b$. This yields a coupling matrix of the form
\begin{equation}
G=\begin{pmatrix}
A_n & -B \\
B^T &-A_m
\end{pmatrix} \;,
\end{equation}
where $A_n$ and $A_m$ represent regular bidirectional ring matrices of link strength $a$ with $k$ and $l$ neighbors, respectively, and $B$ with $B_{ij}=b$ is an $n \times m$ matrix. We impose an equal coupling radius $\kappa$ of the two regular rings, i.e., $\kappa=k/n=l/m$. Thus, together with the unity row sum condition from Eq. (\ref{rowsum}), the weights $a$ and $b$ are fully determined 
\begin{align}
2 \kappa n a - m b &=1 \;, \label{aa} \\
n b - 2 \kappa m a &=1 \;. \label{bb}
\end{align}
Introducing the inhibition ratio, i.e., the ratio of inhibitory to excitatory nodes $\eta=m/n$, Eqs. (\ref{aa}), (\ref{bb}) can be solved as $a=1/(2 \kappa(1-\eta)n)$ and $b=1/(n(1-\eta))$.

\begin{figure}
\includegraphics[height=4cm]{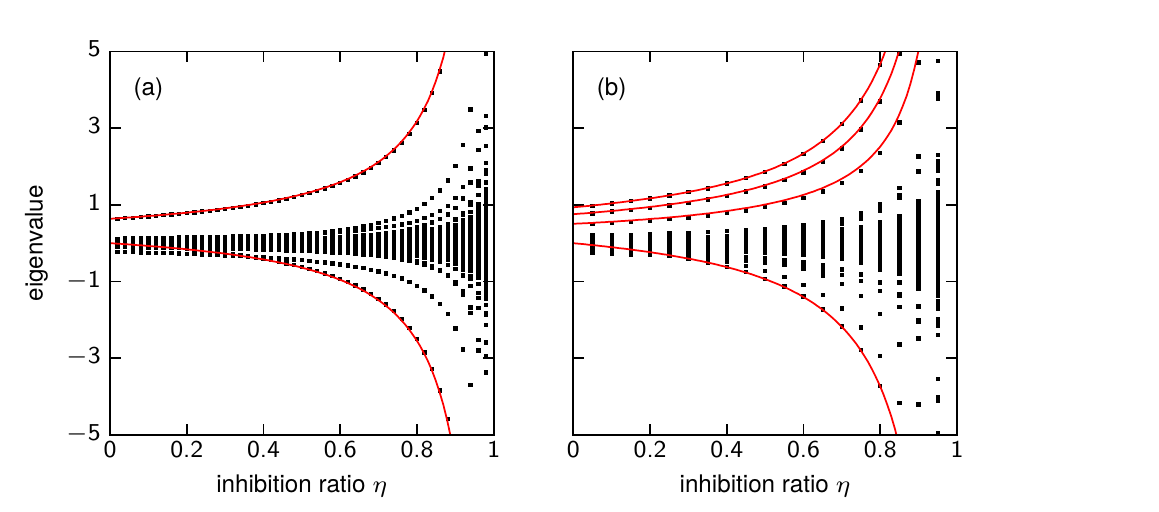}
\caption{(Color online) Eigenvalue spectra (dots) of the coupling matrix as a function of the inhibition ratio $\eta=m/n$ for two different coupling radii (a) $\kappa=0.25$, (b) $\kappa=0.1$. Network size $N=m+n$, with $n=200$ excitatory nodes and $m=\eta n$ inhibitory nodes. The red (grey) lines show the extremal eigenvalues in the large network limit given by Eqs. (\ref{ex1}) and (\ref{ex2}). }
\label{fig_evs}
\end{figure}

Exploiting the block structure of $G$, the eigenvalue spectrum can be constructed explicitly. For a detailed derivation we refer to the Appendix \ref{AppendixA}. The $n+m$ eigenvalues are given by
\begin{equation}
\lambda_j= \frac{1}{2 \kappa n(1-\eta)} \left[ \frac{ \sin(\frac{2 \pi j}{n} (2 \kappa n+1))} {\sin({\pi j}{n}) } -1 \right]
\end{equation}
for $j=1,\ldots,n-1$ and 
\begin{equation}
\lambda_{n+j}= - \frac{1}{2 \kappa n(1-\eta)} \left[ \frac{ \sin(\frac{2 \pi j}{m} (2 \kappa m+1))} {\sin({\pi j}{m}) } -1 \right] \;,
\end{equation}
for $j=1,\ldots, m-1$. The remaining two eigenvalues are $\lambda_{n}=0$ and $\lambda_{n+m}=1$. The latter is present for any matrix with unity row sum. It is called longitudinal eigenvalue and is associated with the dynamics within the synchronization manifold. Hence it can be neglected for the consideration of the stability of a synchronous state \cite{FLU10b,FLU11a}. Thus, in the following description it will be omitted. The spectrum is depicted in Fig. \ref{fig_evs} as a function of the inhibition ratio $\eta$ for two different coupling radii $\kappa$.

In the limit of large networks $n,m \to \infty$ with fixed coupling radius $\kappa$ and inhibition ratio $\eta$, the eigenvalues $\lambda_j, \lambda_{n+j}$ are  well defined for $j \ll n,m$ and yield
\begin{align}
\lambda_{j,\infty} &=\frac{\sin(2 \pi j \kappa)}{2 \pi j \kappa (1-\eta)} \;, \label{ex1} \\ 
\lambda_{n+j,\infty} &= -\frac{\sin(2 \pi j \kappa) \eta }{2 \pi j \kappa (1-\eta)} . \label{ex2}
\end{align}
As they correspond to the extremal eigenvalues depicted by red (black) lines in Fig. \ref{fig_evs}, they will turn out to be of particular importance regarding the stability of synchronization. Due to this fact, it is possible to make statements about the stability for arbitrarily large network sizes within this model. 

Focusing on the large network limit, we note the following observations. The absolute values of the eigenvalues increase with the inhibition ratio $\eta$ and decrease with the coupling radius $\kappa$. For all $\kappa$, there exists a critical inhibition ratio 
\begin{equation}
\eta_c = 1- \frac{\sin(2 \pi  \kappa)  }{2 \pi  \kappa } \;, \label{inhi}
\end{equation}
above which at least one eigenvalue larger than unity exists. Thus, we can confirm and quantify the observation that inhibition leads to a spreading of the eigenvalue spectrum. Further, we note that the distance between the largest and the second largest eigenvalue increases with $\kappa$. For large $\kappa$, this causes the appearance of a gap in the eigenvalue spectrum, which will be relevant for resynchronization at high inhibition ratios. In Fig. \ref{fig_evs} the features mentioned are clearly displayed.

\section{Synchronous Oscillations} \label{sec_sync}
Apart from the network topology, the stability of a synchronous state depends on the dynamics of the network's constituents and the delay. Here, we consider coupled Stuart--Landau oscillators with distributed time delay 
\begin{align}
\dot z_i &= f(z_i)  \nonumber \\
&+ \sigma \sum_{j=1}^N G_{ij} \left[ \int_{0}^\infty  g(t') z_j(t-t') \d t' - z_i(t) \right]  \;, \label{z}
\end{align}
where $f(z)$ is the normal form of a supercritical Hopf bifurcation 
\begin{equation}
f(z)=(\lambda + \mathrm i \omega - |z|^2)z \;,
\end{equation}
$\sigma \in \mathbb R$ is the coupling strength and $G$ denotes the coupling matrix. The delay distribution $g(t)$ is normalized and positive-semidefinite. We consider uniformly distributed delay
\begin{equation}
g_{\text{uni}}(t)=\begin{cases} \frac{1}{\rho} & t \in [\tau-\rho/2,\tau+\rho/2] \\
0 & \text{otherwise}   \end{cases} \;,
\end{equation}
where $\rho$ defines the width of the distribution and $\tau$ is the mean delay time, and Gamma distributions defined by
\begin{equation}
g_{\Gamma}(t) = \frac{(t-\tau_0)^{p-1} l^p  \mathrm e^{- l (t-\tau_0)} }{\Gamma(p)} \Theta (t-\tau_0) \;,
\end{equation}
where $\Gamma(p)$ is the gamma function. Here, $\tau_0$ describes a fixed time offset, which shifts the distribution allowing for coupling without instantaneous parts and the integer $p$ as well as the real parameter $l>0$ characterize the shape of the distribution. The mean delay time $\tau$ is given by $\tau=\int_0^{\infty} g(t) t \d t = \tau_0 + p/l$. 

In the following only the strong kernel, $p=2$, is considered. Generalizations for the weak kernel, i.e., exponential decay ($p=1$) and higher values of $p$, however, can be obtained in a straightforward fashion and exhibit the same qualitative features, since for all Gamma distributions the exponentially decaying tail dominates the behavior.

\begin{figure}
\centering
\includegraphics[height=3.6cm]{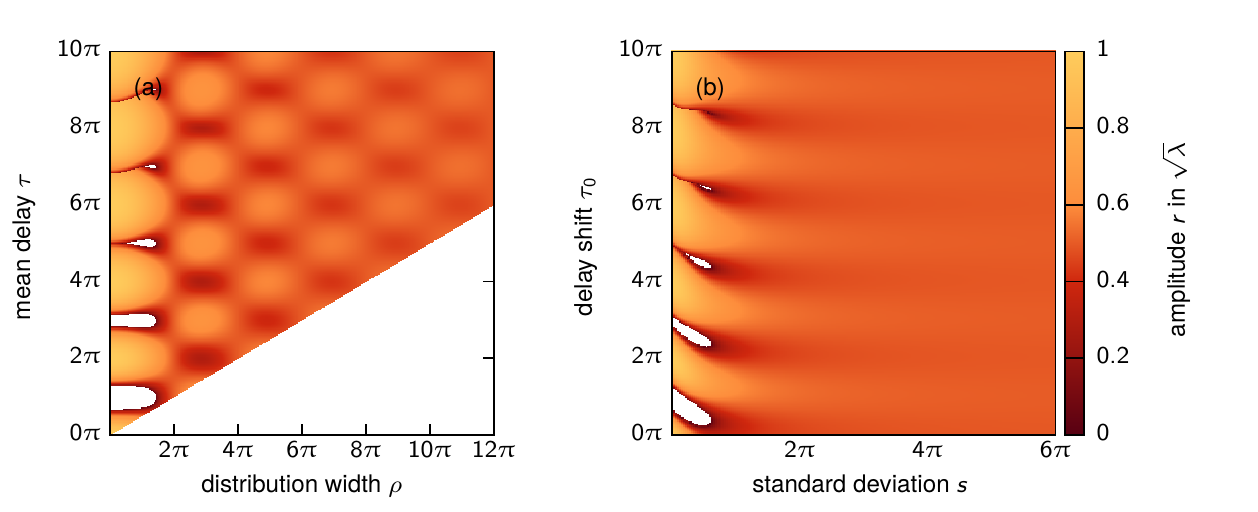}
\caption{(Color online) Amplitude $r$ of the synchronous oscillation for (a) uniformly distributed delay, (b) delay distributed by a $\Gamma_2$-distribution ($p=2$). Parameters: $\lambda=0.12, \sigma=0.09, \omega=1$. Bottom right white region in (a): forbidden because of causality. Other white regions: no solution to Eqs. (\ref{r}), (\ref{o}). }
\label{fig_sol}
\end{figure}

It is convenient to rephrase Eq. (\ref{z}) in polar coordinates using the amplitude $r$ and the phase $\varphi$. With the relation $z=r \mathrm e ^{\mathrm i \varphi}$ one obtains
\begin{align}
\dot r_i(t) &=  (\lambda - r_i^2(t)) r_i(t) + \sigma \sum_{j=1}^N G_{ij} \int_0^\infty g(t') \times \label{rdot}\\
&\left\{  \cos [\varphi_j(t-t') - \varphi_i(t)] r_j(t-t') -r_i(t) \right\} \d t' \;, \nonumber \\
\dot \varphi_i(t) &= \omega + \sigma \sum_{j=1}^N  G_{ij} \int_0^\infty g(t') \times \label{phidot} \\
&  \sin [\varphi_j(t-t')-\varphi_i(t)] \frac{r_j(t-t')}{r_i(t)} \d t' \nonumber \;.
\end{align} 
A synchronous zero-lag oscillation with constant common amplitude $r$ and constant common frequency $\Omega$ implies $r_i=r$ and $\varphi_i=\Omega t$ for all $i$. This yields two coupled equations, which define the synchronous state. For uniformly distributed delay, they are transcendental and read
\begin{align}
r^2 &= \lambda + \sigma \left( \cos (\Omega \tau) \frac{\sin (\Omega \rho) }{\Omega \rho}  -1 \right) \;, \label{r} \\
\Omega &= \omega - \sigma  \sin(\Omega \tau) \frac{\sin (\Omega \rho) }{\Omega \rho}  \;. \label{o}
\end{align}
In the eigenfrequency dominated regime, i.e., $\Omega \simeq \omega$, the existence and the magnitude of the amplitude $r$ follows a periodic structure in both the mean delay $\tau$ and the distribution width $\rho$ (cf. Fig. \ref{fig_sol}). Even (odd) multiples of the eigenperiod $2\pi/\omega$ yield  maximum (minimum) amplitudes, which can be understood intuitively in terms of resonance (antiresonance) phenomena. 

In contrast, for the $\Gamma_2$-distribution ($p=2$) the solution determined by
\begin{align}
r^2 &= \lambda + \sigma \left[ \frac{ (l^2-\Omega^2)l^2 \sin (\Omega \tau_0 + \nu )}{l^2+\Omega^2}    -1 \right] \;, \\
\Omega &= \omega - \sigma \left[ \frac{ (l^2-\Omega^2)l^2 \cos (\Omega \tau_0 - \nu )}{l^2+\Omega^2}    -1 \right] \;,
\end{align}
with 
\begin{equation}
\nu=\arctan \left( \frac{l^2-\Omega^2}{2l \Omega} \right)
\end{equation}
is periodic in the time-offset $\tau_0$, but not periodic in the distribution width, which can be measured by the standard deviation $s=\sqrt{2}/l$ and is inversely proportional to the shape parameter $l$.  
 
 Both distributions reproduce known results for discrete time delay  in the limit of small distribution widths, e.g., the existence of multiple solutions at large delay times \cite{CHO09}, and yield a universal weak coupling solution in the limit of large widths, where $\int g(t') z(t-t') \d t' $ vanishes as a consequence of the Riemann-Lebesgue lemma \cite{BOC49}. The amplitude and frequency in this limit correspond to the solution of uncoupled Stuart--Landau oscillators with eigenfrequency $\omega$ and a shifted bifurcation parameter $\lambda'=\lambda-\sigma$. 
 
 In Fig. \ref{fig_sol}, the amplitude of the synchronous oscillation for (a) a uniform delay kernel and (b) a gamma distributed kernel is displayed as a function of (a) the mean delay time $\tau$ and the distribution width $\rho$ and (b) the time offset $\tau_0$ and the standard deviation $s=\sqrt{2}/l$, respectively. 

\section{Master Stability Function} \label{sec_stable}
The stability of the synchronous solution is determined by the master stability function method \cite{PEC98}, which has recently been extended to systems with distributed time-delay \cite{GJU14}. In this approach, a linear stability analysis of the synchronous solution is performed while the coupling matrix is diagonalized. A variational equation defines the Floquet exponents $\Lambda$, which determine the stability.

\begin{figure}
\centering
\includegraphics[width=0.63\textwidth]{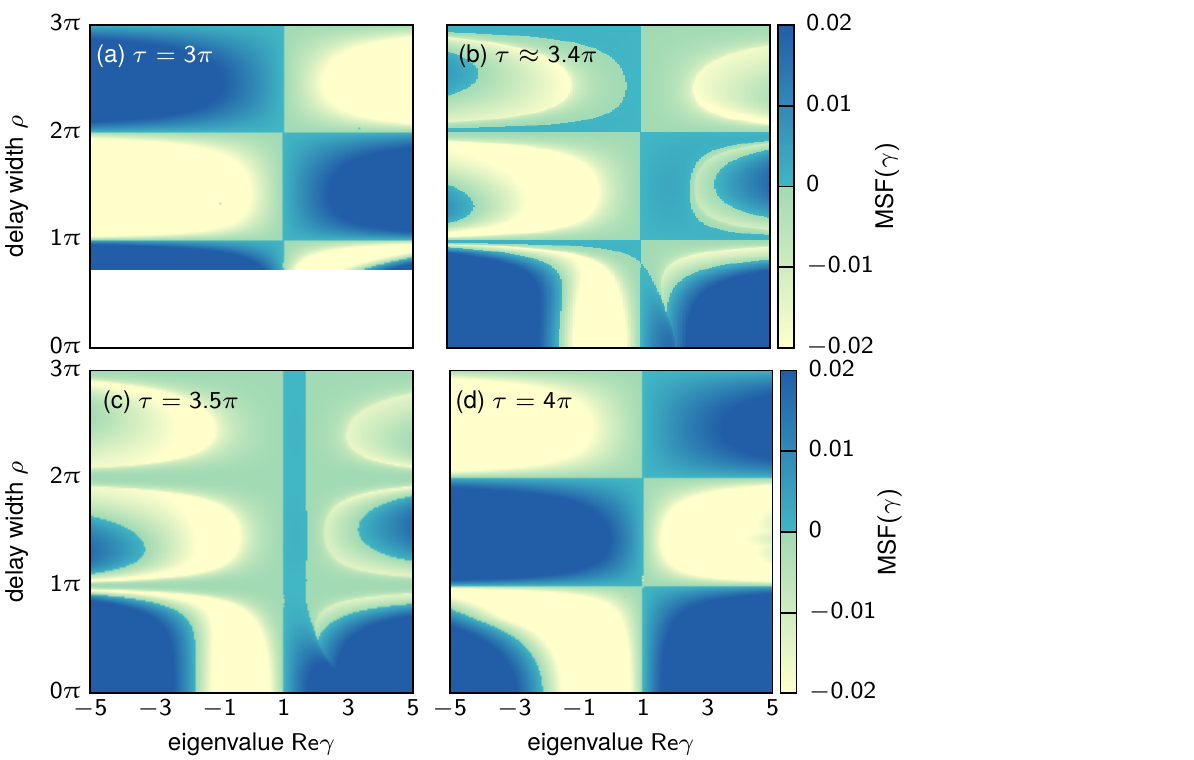}
\caption{(Color online) Master stability function (MSF) in the $(\re \gamma,\rho)$-plane for uniformly distributed delay of width $\rho$ with different mean delay times $\tau$; $\gamma$ corresponds to the eigenvalues of the coupling matrix. Parameters: $\lambda=0.12, \omega=1, \sigma=0.09$. White region in (a): no solution ($r \in \mathbb R$) to Eqs. (\ref{r}), (\ref{o}).}
\label{fig_msf_rect}
\end{figure}

As we consider a periodic orbit, a description in radial and phase coordinates reflects the symmetry of the problem and yields time-independent Jacobi matrices, which simplifies the stability analysis drastically. However, in this coordinate system, the coupling term is non-linear, which requires a technical modification of the approach given in Ref. \cite{GJU14}. The case considered here in Eqs. (\ref{rdot}), (\ref{phidot}) is of the general form 
\begin{equation}
\dot {\ve x}_i = \ve f (\ve x_i) + \sum_{j=1}^N G_{ij} \int_0^{\infty} g(t') \ve h(\ve x_i, \ve x_j(t-t')) \d t' \;,
\end{equation}
where the coupling scheme $\ve h(\ve x_i, \ve x_j(t-t))$ reduces to a matrix vector multiplication for linear coupling as in Ref. \cite{GJU14}. The linear stability of the synchronous solution is determined by the following variational equation
\begin{equation}
\delta \dot{\ve x} = J(t) \delta \ve x + \gamma \int_0^{\infty} g(t') A(t,t') \delta \ve x(t-t') \d t' \;,
\end{equation}
where the matrices $J(t)$ and $A(t,t')$ are calculated as
\begin{align}
J_{\alpha \beta}(t) &= \left. \frac{\partial f_\alpha(\ve x)}{\partial x_\beta} \right\lvert_{\ve x_s} \\
& + \int_0^{\infty} g(t') \left. \frac{\partial h_\alpha(\ve x, \ve x(t-t'))}{\partial x_\beta} \right\lvert_{\ve x_s} \d t' \nonumber \;,\\
A_{\alpha \beta}(t,t') &= \left. \frac{\partial h_\alpha(\ve x, \ve x(t-t')}{\partial x_\beta(t-t')} \right\lvert_{\ve x_s}
\end{align}
and
%
$\gamma$ corresponds to the eigenvalues of the coupling matrix that emerge from the diagonalization of $G$, and $\ve x_s$ represents the synchronous solution. 

In the case studied here, the variational state vector is defined by $\delta \ve x=(\delta r, \delta \varphi)$ and due to the choice of the coordinate system, the matrices $J(t)$ and $A(t,t')$ do not explicitly depend upon time $t$, i.e., $J(t)=J$ and $A(t,t')=A(t')$. They read
\begin{align}
J &= \begin{pmatrix}
-2 r^2 - \sigma C(\Omega) & -\sigma S(\Omega) \\
\sigma S(\Omega) & -\sigma C(\Omega)
\end{pmatrix} \;, \\[0.2cm]
A(t') &= \sigma \begin{pmatrix}
\cos \Omega t' & \sin \Omega t' \\
- \sin \Omega t' & \cos \Omega t' 
\end{pmatrix} \;,
\end{align}
with the abbreviations 
\begin{align}
C(\Omega ) &=\int_{0}^\infty g(t') \cos (\Omega t') \d t' \;, \\
S(\Omega ) &=\int_{0}^\infty g(t') \sin (\Omega t') \d t \; .
\end{align}
Applying the ansatz $\delta \ve x \propto e^{\Lambda t}$ yields a generalized characteristic equation for the Floquet exponents $\Lambda$
\begin{equation}
\det[ J - \gamma \int_{0}^{\infty} g(t') A(t') \mathrm e ^{-\Lambda t'} \d t' - \Lambda \mathrm I ] =0 \;, \label{det}
\end{equation}
where $\mathrm I$ is the identity matrix. The maximum real part of the Floquet exponents determines the stability. Evaluated as a function of the eigenvalue $\gamma$, it is referred to as master stability function (MSF)
\begin{equation}
\text{MSF}(\gamma)=\max_\Lambda \re \Lambda(\gamma) \; .
\end{equation}
  When all eigenvalues $\gamma$ of a given coupling matrix $G$ yield a negative value of the master stability function, the synchronous state is stable for this network. Since for the network model considered here the eigenvalues are real, the MSF is evaluated only for real $\gamma$.
 
The generalized characteristic equation (\ref{det}) is equivalent to determining the roots of an exponential polynomial. Exploiting the analyticity of these functions yields an elegant procedure to determine the number of roots with positive real parts, which answers the question of stability without calculating the Floquet exponents explicitly. For uniformly distributed delay this procedure is explained in Appendix \ref{AppendixB}.

The MSF for uniformly distributed delay inherits some of the characteristic structures from the solution itself (cf. Fig. \ref{fig_msf_rect}). Its qualitative appearance is periodic in the mean delay and the distribution width. Evaluated in the $(\re \gamma, \rho)$-plane, for mean delay times equal to multiples of half the eigenperiod $2 \pi/\omega$, i.e., $\tau=\pi,2\pi,3\pi,\ldots$ (Fig. \ref{fig_msf_rect}a, d) the periodicity in $\rho$ resembles the checkerboard structure of resonances and antiresonances, which determines the amplitude of the synchronous solution (cf. Fig. \ref{fig_sol}). In particular, antiresonant regions are characterized by the existence of a stability island only for $\re \gamma>1$, a condition that can not be matched by all eigenvalues of a coupling matrix of the type considered here and thus corresponds to unstable synchronization.

\begin{figure}
\includegraphics[width=\columnwidth]{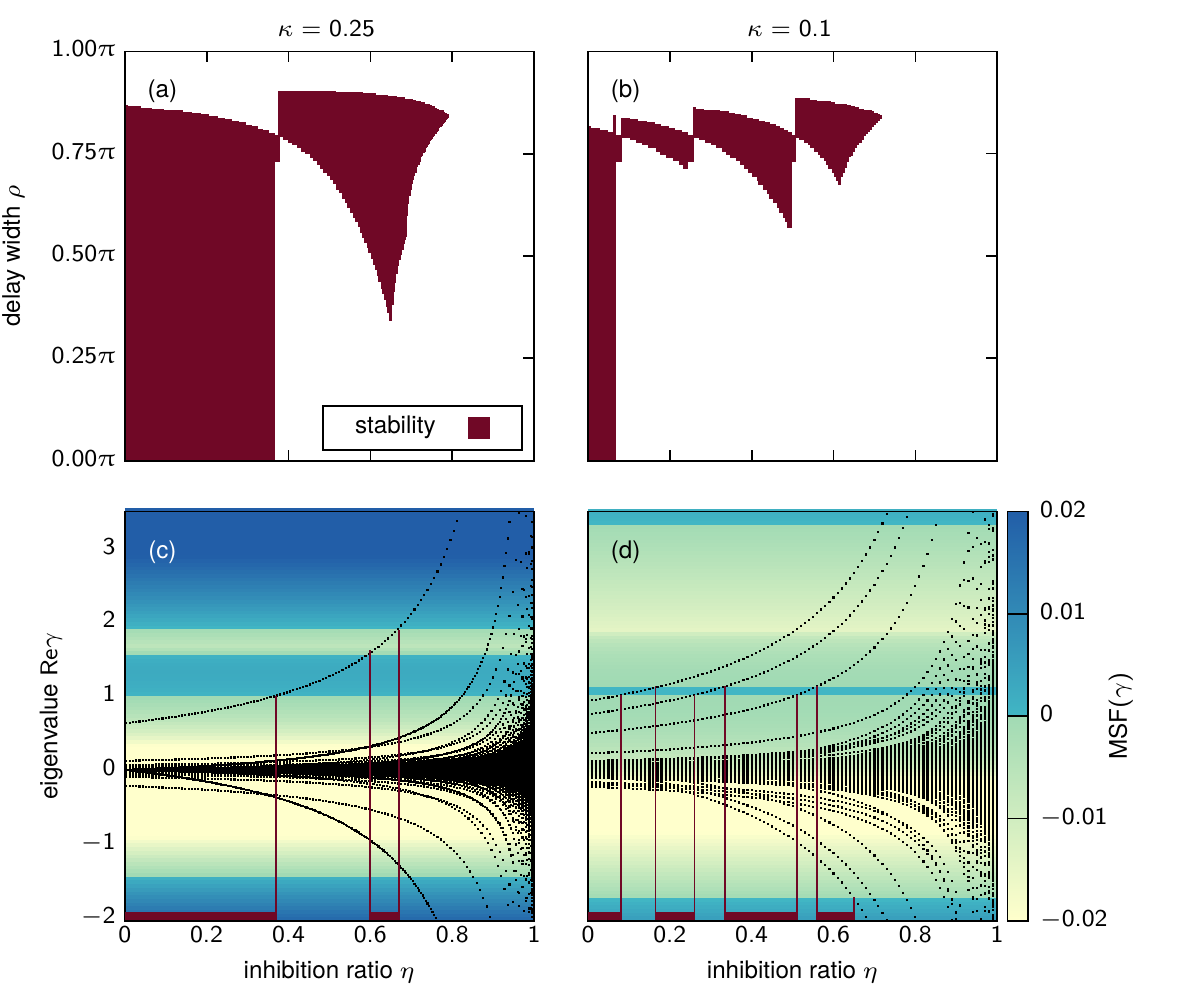}
\caption{(Color online) Stable synchronization as a function of the inhibition ratio $\eta$ and the delay width $\rho$ for uniformly distributed delay in a network with coupling radius (a)  $\kappa=0.25$, (b) $\kappa=0.1$. Eigenvalue spectra (dots) and master stability function (color code (grey scale)) with (c) coupling radius ratio $\kappa=0.25$ and delay distribution width $\rho=0.5\pi$ and (d) $\kappa=0.1, \rho=0.75 \pi$. Network size $N=n+m$ with $n=2000$ excitatory and $m=\eta n$ inhibitory nodes. Other parameters as in Fig. \ref{fig_msf_rect} (b). Vertical lines in (c), (d) denote synchronization-desynchronization transitions.}
\label{fig_desync}
\end{figure}

In the resonant regions the stability island (light color) contains the interval $[-1,1]$. Hence, networks below the critical inhibition ratio given in Eq. (\ref{inhi}) exhibit stable synchronous oscillations. The same applies to mean delays exactly between multiples of half the eigenperiod, i.e., multiples of a quarter eigenperiod (cf. Fig. \ref{fig_msf_rect}c). Here, the unit interval is contained within the stability regions for any $\rho$. Additionally, a second stability island appears above a certain delay distribution width $\rho$. The second stability island can induce stable synchronization for networks with an inhibition ratio above the critical one. This is the case if the eigenvalues outside the unit interval happen to lie inside the additional stability island. 

For any mean delay different from the special cases mentioned above, the MSF shows a very rich structure (Fig. \ref{fig_msf_rect}b). Depending on $\rho$ a single connected or two disconnected stability regions of various size can occur. Further, the gap between the two islands can lie left or right to the line $\re \gamma=1$. Note that similar behavior has been found in case of a discrete delay and different local dynamics \cite{KEA12}. The stability gap in the master stability function immediately implies several interesting consequences for the stability of synchronous solutions in inhibitory networks, which are discussed in detail in the following. 

\section{Synchronization-Desynchronization Transitions} \label{sec_trans}

As the interval $\rho \in [0, \pi]$ in Fig. \ref{fig_msf_rect}b covers all qualitatively different regimes, further analysis is restricted to this interval. In Fig. \ref{fig_desync} the stability of synchronous solutions is evaluated for the concrete network type proposed here in the parameter setting that yields the MSF of Fig. \ref{fig_msf_rect}b. The stability is evaluated as a function of the inhibition ratio $\eta$ and the distribution width $\rho$ of the delay for two different coupling radii $\kappa$. The result elucidates in detail the complex interplay between distributed delay and the excitatory-inhibitory node balance.

For the higher coupling radius $\kappa=0.25$ (cf. Fig. \ref{fig_desync}a), the stability of synchronization has a strong structural agreement with the MSF itself (cf. Fig. \ref{fig_msf_rect}b). Both can be categorized into five different $\rho$-regimes to get a better understanding of how their features are related.

(i) For small distribution widths, i.e., $\rho<0.37\pi$ only one stability island of the MSF exists. It contains the interval $[-1,1]$ and is bounded by $\re \gamma=1$. Thus, all networks up to the critical inhibition ratio exhibit stable synchronous solutions and above the critical inhibition ratio, the stability collapses and the network desynchronizes. 

(ii) For larger $\rho$, i.e., $0.37\pi<\rho<0.75\pi$ a second stability island appears in the MSF for $\re \gamma>1$ and its extension grows with $\rho$. This causes a resynchronization at relatively high inhibition ratios, when the largest eigenvalue increasing with $\eta$ reenters the stability region. For $\rho=0.5\pi$ this is depicted in Fig. \ref{fig_desync}c. Since for stability all eigenvalues of the coupling matrix are required to lie inside the stability regions, the gap of the eigenvalue spectrum of $G$ is crucial. It must contain the stability gap of the MSF in order to induce resynchronization above the critical inhibition ratio. 

(iii) In a narrow $\rho$-range around $\rho=0.8\pi$, the two stability islands of the MSF almost touch each other. Here, synchronization is stable up to extremely high inhibitory ratios of about $\eta \approx 0.75$. 

(iv) For even larger $\rho$, i.e., $0.8 \pi < \rho < 0.9 \pi$, the stability gap of the MSF lies within the interval $[-1,1]$. Hence, synchronization is suppressed for comparably small inhibition ratios when the maximum eigenvalue is no longer contained inside the stability region. On the other hand, resynchronization occurs at lower inhibition ratios. In an extreme case, this leads to the situation that synchronization is unstable up to a certain inhibition ratio, which is just the inverse behavior of what occurs for small $\rho$ in case (i). 

(v) Finally, synchronization is unstable for any network, if the stability island on the left is too small, which is the case for $0.9 \pi < \rho < \pi$. In particular if the eigenvalue $\lambda_m=0$, which is independent of $\kappa$ and $\eta$, is no longer contained inside the stability region.

For networks with lower coupling radii, e.g. $\kappa=0.1$, the eigenvalue spectrum has no pronounced gap below the largest eigenvalue. Thus, several of the largest and smallest eigenvalues leave and reenter the gapped stability regions of the MSF when the inhibition ratio is increased. This leads to qualitative deviations from the case $\kappa=0.25$ as displayed in Fig. \ref{fig_desync}b. Apart from the decreased critical inhibition ratio defined in Eq.~(\ref{inhi}), the most striking difference appears for intermediate values of $\rho$, where instead of a single resynchronization, multiple synchronization-desynchronization transitions appear for increasing $\eta$. Each of them is caused by a different eigenvalue $\lambda$. This mechanism is visualized in Fig. \ref{fig_desync}d.

The last two examples illustrate how crucial the delay distribution width is for the stability of synchronization in a network with inhibitory nodes. Depending on the distribution width, the inhibition ratio can induce a single synchronization transition, a single desynchronization transition or even multiple synchronization-desynchronization transitions. For small coupling radii the effect of inhibition is enhanced, and multiple transitions are more likely to occur. 

\begin{figure}
\includegraphics[width=0.55\textwidth]{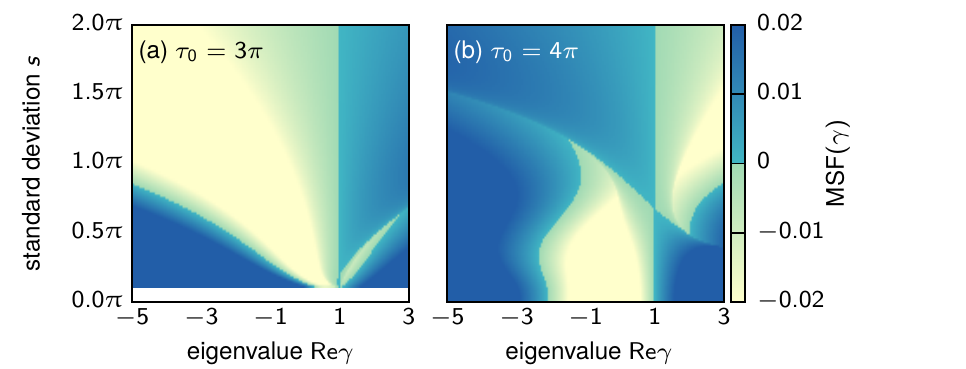}
\caption{(Color online) Master stability function for a $\Gamma_2$ delay distribution with standard deviation $s$ in the $(\re \gamma,s)$-plane for two different time-offsets (a) $\tau_0=3\pi$, (b) $\tau_0=4\pi$. Other parameters as in Fig. \ref{fig_msf_rect}.}
\label{fig_msf_gam}
\end{figure}

We emphasize that for the network type chosen here, every transition can be predicted analytically from the eigenvalue spectrum in the continuum limit once the MSF is known. Thus, we can conclude universal statements from the MSFs by categorizing them into the five distinct regimes listed above. This procedure is independent of the specific form of the distribution. 

For instance, the MSF for the $\Gamma_2$-distribution can be categorized by the same means. In the $(\re \gamma, s)$-plane of the topological eigenvalue $\gamma$ and the standard deviation $s$ of the delay distribution, evaluated at different time-offsets $\tau_0$ it exhibits similar structures and regularities like e.g., the stability gap and the particular importance of the line $\re \gamma=1$ (cf. Fig.~\ref{fig_msf_gam}). For suitable values of $\tau_0$, e.g., $\tau_0=4\pi$, the MSF shows qualitatively the same features as the MSF for uniformly distributed delay from Fig.~\ref{fig_msf_rect}b. Thus, it will generate a similar stability diagram as depicted in Fig. \ref{fig_desync}. This implies two useful consequences. First, the rather artificial uniform distribution can be used to approximate the more realistic Gamma distribution for suitable choices of the parameters. Second, the stability diagram in Fig. \ref{fig_desync} possesses a certain universality in the sense that for this type of model qualitatively similar regimes are expected.

\section{Robustness of Results} \label{sec_asym}
The network model considered here has the advantage that it can be treated analytically due to its high degree of symmetry. However, real-world networks will typically possess a certain degree of asymmetry. Thus, it is interesting to investigate whether the qualitative nature of the synchronization--desynchronization transitions observed for this specific network model is robust against asymmetric perturbations of the network topology. 

In order to address this question, we propose a rewiring algorithm which breaks the network symmetry and investigate the synchronization--desynchronization transitions for those perturbed networks.

For an arbitrary asymmetric network the stability of synchronization can no longer be calculated analytically since the topological eigenvalues of the coupling matrix are not known in closed form and have to be computed numerically. They are complex and thus have to be compared to the master stability function evaluated in the $(\re \gamma, \im \gamma)$-plane.

We choose a rewiring algorithm that leaves the number of inhibitory and excitatory nodes constant such that the inhibition ratio $\eta$ remains a characteristic network parameter. To this end we pick a node randomly and rewire one of its outgoing links to another randomly chosen node, with duplicated links forbidden.
The weight of the new link is chosen according to the four connection classes, i.e., excitatory-inhibitory (exc.-inh.) links of weight $b$, exc.-exc. links of weight $a$, inh.-inh. links of weight $-a$, and inh.-exc. links of weight $-b$. To ensure synchronizability the row sums are normalized after the rewiring process. 

Note that in networks with inhibition the normalization procedure is ambiguous and a normalized row sum in principle still allows for inputs with arbitrarily large absolute values as long as excitation and inhibition compensate each other. To avoid these unphysical situations we impose the condition that in addition the row sum of absolute values is kept constant. This ensures that the eigenvalue spectrum of the coupling matrix is bounded by the same Gershgorin discs before as well as after the rewiring procedure.

Cases where the row sum is no longer positive are excluded by the algorithm since the row sum cannot be normalized in a straightforward fashion.

By rewiring repeatedly, the total number of inhibitory and excitatory nodes as well as the total number of excitatory and inhibitory links is kept constant, but the topology is steadily randomized under these constraints. The global coupling between nodes of different categories acquires defects, while the regular ring networks defining the topology of nodes of the same category transform into small-world networks \cite{WAT98}. In the limit of infinitely many rewirings, the signatures of the regular rings and of the global coupling are lost and the network evolves into a random network with a uniform wiring probability per connection category. 

\begin{figure}
\centering
\includegraphics{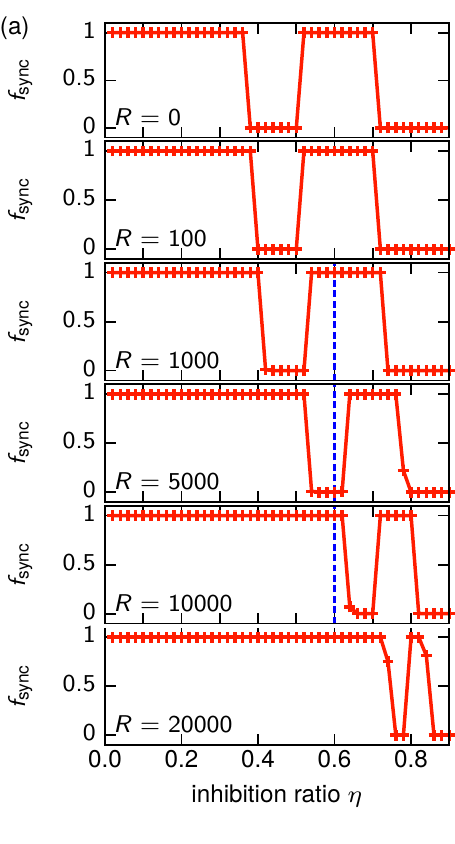} \includegraphics{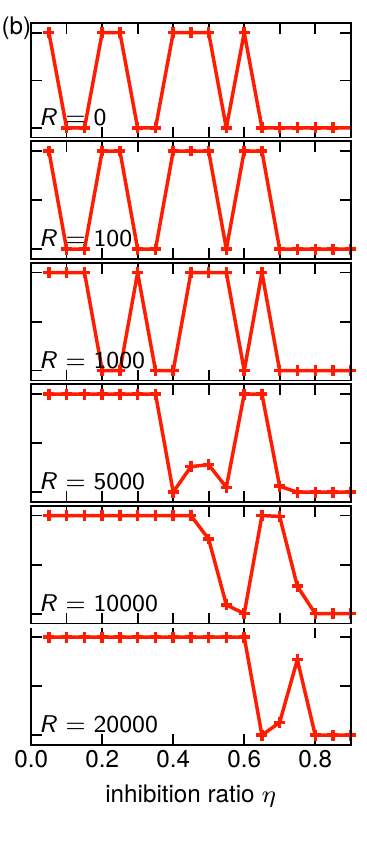}
\caption{(Color online) Fraction of stable synchronization for asymmetric networks emerging from the rewiring process described in 
Section \ref{sec_asym} as a function of the inhibition ratio $\eta$ for increasing numbers of repeated rewirings 
$R$ from top to bottom. The blue (black) dashed line marks parameters which are considered in Fig. \ref{fig_msf_ev}. Parameters: $100$ realizations, network size $N=n+m$ with $n=200$ excitatory and $m=\eta n$ 
inhibitory nodes, (a) $\rho=0.7 \pi$, $\kappa=0.25$, (b) $\rho=0.75 \pi$, $\kappa=0.1$, other parameters as in Fig. \ref{fig_msf_rect}b. }
\label{fig_asym}
\end{figure}

In Fig.~\ref{fig_asym} we show the fraction of synchronization $f_\text{sync}$ as a function of the inhibition rate $\eta$ for the unperturbed symmetric network ($R=0$) as well as for perturbed networks generated by more and more rewiring iterations $R$. We consider two different coupling radii $\kappa$ and delay distribution widths $\rho$ that show non trivial synchronization--desynchronization transitions for the unperturbed network in Fig.~\ref{fig_desync}, i.e., $\kappa=0.25$, $\rho=0.75 \pi$ in Fig.~\ref{fig_asym}a and $\kappa=0.1$, $\rho=0.7\pi$ in Fig.~\ref{fig_asym}b. The qualitative nature of the transitions is preserved for rewiring iterations of up to $R=20000$ in (a), and, up to $R=1000$ in (b), respectively. At $R=1000$ and for coupling radius $\kappa=0.1$ considered in (b), the parts of the network connecting nodes of the same category are no longer well described by regular rings, but exhibit the characteristics of a small-world network namely small shortest pathlengths and high clustering. It is interesting that for these networks the synchronization-desynchronization transition is qualitatively the same as compared to the fully symmetric network. 

However, the transitions occur at higher inhibition ratios for rewired networks compared to unperturbed networks. For a very large number of rewirings ($R=10000,20000$) the synchronization--desynchronization transition changes qualitatively for $\kappa=0.1$ (Fig. \ref{fig_asym}b) and synchronization is stable up to much larger inhibition ratios for both $\kappa=0.25$ and $\kappa=0.1$. 

\begin{figure}
\includegraphics{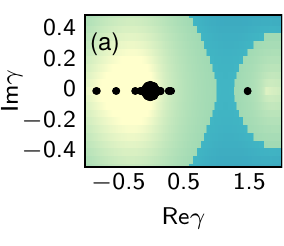} \includegraphics{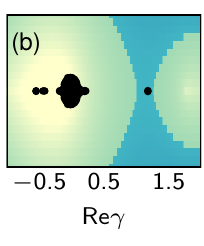}  \includegraphics{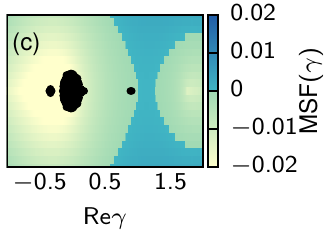}
\caption{(Color online) Stability of synchronization in asymmetrically perturbed networks. Color (gray) shading: master stability function, black dots: eigenvalue spectrum for a network generated by (a) $R=1000$, (b) $R=5000$, and (c) $R=10000$ rewirings. Parameters: $\eta=0.6$, $\kappa=0.25$ (see Fig.~\ref{fig_asym}a blue (black) dashed line), $n=200$, $\rho=0.7 \pi$, others as in Fig.~\ref{fig_msf_rect}b. }
\label{fig_msf_ev}
\end{figure}

This behavior can be understood by considering how the eigenvalue spectrum transforms under the rewiring process. In Fig. \ref{fig_msf_ev}, we show the master stability function compared to the eigenvalue spectra of networks generated by an increasing number of rewirings. The following trends are visible: For an increased number of rewirings, the imaginary parts of the eigenvalues increase due to the emergence of asymmetry as expected. In addition, the extremal eigenvalues are dragged towards the origin and the spectrum is contracted to a spherical distribution of comparably small radius, which is characteristic for random networks \cite{JAL11a}. The contraction explains why the synchronization-desynchronization transitions occur for higher inhibition ratios compared to the unperturbed networks, since the shrinking of the maximum real part counteracts the spreading of the eigenvalues along the real axis caused by increasing inhibition ratios (see Fig. \ref{fig_desync}c). 

Since the multiple synchronization--desynchronization transitions occur due to the gap structure of the eigenvalue spectrum for the symmetric model, we infer that they do not appear for strongly randomized networks whose eigenvalue spectra show less pronounced gaps. This explains why we observe a qualitative change of the synchronization--desynchronization transition in Fig. \ref{fig_asym}b from $R=1000$ to $R=5000$.

We conclude that the analytic results obtained for the symmetric network model are transferable to networks with 
asymmetric perturbations. However, qualitative deviations from the latter are possible if the information about the 
original network structure is destroyed by sufficiently large perturbations.

\section{Conclusion} \label{sec_conc}
In this paper we have introduced a dynamic network model, where the influence of the excitatory-inhibitory node balance could be directly tracked in the eigenvalue spectrum of the coupling matrix. For this model we could thus confirm and extend previous results that increasing inhibition ratios lead to a spreading of the eigenvalue spectrum which likely causes desynchronization. 

The existence of synchronous oscillations has been investigated in a network of coupled Stuart-Landau oscillators with distributed delay. For uniformly distributed delay, the amplitude of oscillations depends in a periodically modulated way on the distribution width and the mean delay, which we can understand as a consequence of resonance and antiresonance phenomena.

This characterization into resonant and antiresonant regimes could also be applied to classify the structure of the master stability function (MSF), which determines the stability of synchronous oscillations. Here, we have identified regions of resonance and antiresonance determined by the position of the stability island of the MSF. However, for mean delays incommensurable with half the eigenperiod stability gaps are observed. The resulting consequences for the stability of synchronous oscillations in networks with inhibitory nodes were analyzed in detail and the combined effects of distributed delay and the inhibition ratio was investigated. We have found that both the delay distribution width as well as the inhibition ratio can induce synchronization-desynchronization transitions. For small delay distribution widths, strong inhibition causes desynchronization. This result is in agreement with other works, e.g., by Battaglia et al.\cite{Bat07}, where sufficiently strong long-range excitations cause full synchronization.  As a tendency, we found that distributed delay enhances the possibility of stable synchronization for increasing inhibition ratios. 

The robustness of our results against asymmetric perturbations of the network topology is confirmed numerically. One main result is that the synchronization-desynchronization transitions do not change qualitatively when random rewiring of the links introduces a small-world structure for the connections between nodes of the same category. This regime is of particular interest since it is known that neural networks exhibit small-world character \cite{STA04d,BAS06a}.

This work indicates the relevance of distributed delay beyond the stabilization of fixed points and sheds some light on its interesting effects on synchronization in non-linear systems.

\section{Acknowledgments}
This work was supported by DFG in the framework of SFB 910.
\appendix
\section{Eigenvalues of the Coupling Matrix}  \label{AppendixA}
To calculate the eigenvalues of 
\begin{equation}
G=\begin{pmatrix}
A_n & -B \\
B^T &-A_m
\end{pmatrix} \;,
\end{equation}
where $A_n$ and $A_m$ are regular bidirectional ring $(n \times n$) and $(m\times m)$ matrices, respectively, and $B$ is an $n\times m$ matrix with entry $b$ at every position, we make the following ansatz for the $(n+m)$-dimensional eigenvector $\ve v = (\ve x, \ve y)$. Here, $\ve x$ is an $n$-dimensional vector and $\ve y$ is an $m$-dimensional vector. The eigenvalue problem then reads
\begin{align}
A_n \ve x - B \ve y &= \lambda \ve x \;, \\
B^T \ve x - A_m \ve y &= \lambda \ve y \;.
\end{align}
Setting $\ve y=\ve 0$, we obtain 
\begin{align}
A_n \ve x &= \lambda \ve x \;, \\
B^T \ve x &= \ve 0\;. \label{B}
\end{align}
As $A_n$ is a circulant matrix, its eigenvectors are known and the $l$-th entry of the $j$-th eigenvector is given by
\begin{equation}
x_l^{(j)}= \exp \left( \frac{ 2\pi \mathrm i j (l-1)}{n} \right) \;, \quad l,j=1,\ldots,n \;. \label{A6}
\end{equation}
Since $B$ has the same entry at every position, Eq. (\ref{B}) yields $b \sum_{l=1}^n x_l =0$, which is fulfilled for every $\ve x^{(j)}$ defined in Eq. (\ref{A6}) with $j=1,\ldots,n-1$. Hence, the first $n-1$ eigenvectors are found and the corresponding eigenvalues can be calculated from the general formulas known for circulant matrices. With the ansatz $\ve x=0$, further $m-1$ eigenvectors can be constructed analogously. The eigenvector $(1,1,\ldots,1)$ yields the eigenvalue $1$ due to the unit row sum condition and the last eigenvector can be calculated explicitly with the ansatz $x_l=x$, $y_l=1$. A straightforward calculation yields $x=m/n$ and the corresponding eigenvalue is again calculated explicitly.

\section{Zeros of the Exponential Polynomial } \label{AppendixB}
For uniformly distributed delay the problem of solving Eq. \F{det} is essentially a problem of determining the roots of an exponential polynomial with frequencies $\tau+\rho$ and $\tau-\rho$. To see this, one has to multiply Eq. \F{det} with $(\Lambda^2+\Omega^2)$. Then, problem Eq. \F{det} can be written in the following form
\begin{equation}
P(\Lambda)=0 \;,
\end{equation}
where 
\begin{equation}
P(\Lambda)=\Lambda^4+\sum_{i=0}^3 \sum_{j,k=0}^2 a_{ijk} \, \Lambda^i \mathrm e^{- \Lambda (\tau-\rho) j} \mathrm e^{- \Lambda (\tau+\rho) k} 
\end{equation}
with real coefficients $a_{ijk}$, given by
\begin{align*}
a_{000} &=c_0 \Omega^2 \;, \\ 
a_{010} &= \frac{2 \gamma  \sigma \Omega}{\rho} \left[ \sin (   \Omega \rho + \Omega \tau ) (-2   r^2-2 g_1 \sigma) \right. \\
& \left. \quad +2 g_2 \sigma \cos (\Omega \rho  + \Omega \tau )   \right] \;, \\
a_{001} &=  \frac{2 \gamma  \sigma \Omega}{\rho} \left[ \sin (  \Omega \rho - \Omega \tau ) (2  r^2+2 g_1 \sigma)  \right. \\
& \left. \quad-2 g_2 \sigma \cos (\Omega \rho  - \Omega \tau )   \right] \;, \\
a_{011} &= -\frac{ \gamma^2 \sigma^2 }{2 \rho^2} \cos (2 \rho  \Omega ) \;,\\
a_{020} &= a_{002} =\frac{\gamma^2 \sigma^2}{4 \rho^2} \;,\\
a_{100} &= 2 \Omega^2 (g_1 \sigma + r^2)\;, \\
a_{110} &= \frac{ \gamma  \sigma }{2 \rho} \left[ \cos ( \Omega \rho + \Omega \tau ) (-2  r^2-2 g_1 \sigma) \right. \\
& \quad \left. -2 g_2 \sigma  \sin ( \Omega \rho  + \Omega \tau )  - \Omega \sin( \Omega \rho +\Omega \tau) \right]\;, \\
a_{101} &=  -\frac{ \gamma  \sigma }{4 \rho} \left[ \cos (  \Omega \rho - \Omega \tau ) (-2  r^2-2 g_1 \sigma)  \right. \\
& \left. \quad -2 g_2 \sigma  \sin ( \Omega \rho  - \Omega \tau )  + \Omega \sin(\Omega \rho -\Omega \tau) \right]  \;,\\
a_{200} &= c_0 +\Omega^2 \;, \\
a_{210} &= - \frac{ \sigma \gamma}{4 \rho} \cos (\Omega\rho+\Omega \tau)\;, \\
a_{201} &= \frac{ \sigma \gamma}{4 \rho} \cos (\Omega\rho -\Omega \tau)\;, \\
a_{300} &= 2 \sigma g_1 + 2 r^2 \;,
\end{align*}
where 
\begin{align*}
c_0 &= g_1^2 \sigma^2 +2  g_1 \sigma r^2+g_2^2 \sigma^2 \;, \\
g_1 &=\frac{\sin (\Omega \rho)}{\Omega \rho} \cos (\Omega \tau) \;, \\
g_2 &=-\frac{\sin (\Omega \rho)}{\Omega \rho} \sin (\Omega \tau) \;.
\end{align*}
The synchronous state is stable if $P(\Lambda)$ has no roots with positive real part. This question can be answered using fundamental results of complex analysis since $P(\Lambda)$ is an analytic function, which decays exponentially with $\re \Lambda \to \infty$. Following the work by Habets \cite{HAB92} with a slight modification regarding the treatment of the roots on the imaginary axis $\pm \mathrm i \Omega$, we find that the number of roots $N$ with positive real parts can be calculated as
\begin{align}
N &= 1 - \frac{1}{2 \pi } \left( \int_{- B}^{- \Omega -  \epsilon}\frac{P'( \mathrm i \omega )}{P(\mathrm i \omega)} \d  \omega  + \int_{-  \Omega +  \epsilon}^{ \Omega -  \epsilon} \frac{P'(\mathrm i \omega)}{P(\mathrm i \omega)} \d  \omega  \right. \nonumber \\
& \left. \quad +\int_{  \Omega + \epsilon}^{ B} \frac{P'(\mathrm i \omega)}{P(\mathrm i \omega)} \d \omega \right)  \;,
\end{align}
where
\begin{equation}
B=\sqrt \alpha_{\text{max}} + \max(1,\alpha_3)
\end{equation} 
with $\alpha_i=\sum_{jk} |a_{ijk}|$ and $\alpha_\text{max} = \max 
\{ \alpha_i | i=0,1,2 \}$ and $\epsilon \to 0$. This reduces the multidimensional root search to the calculation of a single integral of a real-valued function on a finite interval, which is an enormous simplification of the problem. Additionally, this method is very robust against numerical errors and produces a reliable boundary of the stability islands of the MSF. 

%
\end{document}